\documentclass[aps,prd,twocolumn,showpacs,superscriptaddress,floatfix]{revtex4-1}

\usepackage{graphicx}
\usepackage{amsmath,amssymb,bm}
\usepackage{multirow}
\usepackage{float}
\usepackage{hyperref}
\usepackage{booktabs}
\providecommand{\mnras}{Mon. Not. R. Astron. Soc.}

\hypersetup{colorlinks=true,linkcolor=blue,citecolor=blue,urlcolor=blue}
\newcommand{\orcid}[1]{%
  \href{https://orcid.org/#1}{%
    \includegraphics[height=2ex, keepaspectratio]{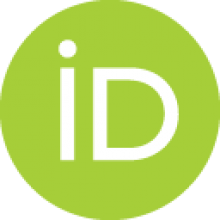}%
  }%
}
\newcommand{\DM}{\mathrm{DM}}
\newcommand{\DME}{\mathrm{DM}_{\rm E}}
\newcommand{\DMIGM}{\mathrm{DM}_{\rm IGM}}
\newcommand{\Om}{\Omega_{\rm m}}
\newcommand{\Ob}{\Omega_{\rm b}}
\newcommand{\fd}{f_{\rm d}}

\begin{document}

\title{Interpretable attention-based neural posterior estimation for fast radio burst cosmology}

\author{Tian-Yang Sun\orcid{0009-0002-5109-6420}}
\email{suntianyang@stumail.neu.edu.cn}
\affiliation{Liaoning Key Laboratory of Cosmology and Astrophysics, College of Sciences, Northeastern University, Shenyang 110819, China}

\author{Ji-Guo Zhang\orcid{0009-0004-5607-9181}}
\email{zhangjiguo@stumail.neu.edu.cn}
\affiliation{Liaoning Key Laboratory of Cosmology and Astrophysics, College of Sciences, Northeastern University, Shenyang 110819, China}

\author{Wan-Peng Sun\orcid{0000-0001-5950-7170}}
\email{sunwp@mails.neu.edu.cn}
\affiliation{Liaoning Key Laboratory of Cosmology and Astrophysics, College of Sciences, Northeastern University, Shenyang 110819, China}

\author{Jing-Fei Zhang\orcid{0000-0002-3512-2804}}
\email{jfzhang@mail.neu.edu.cn}
\affiliation{Liaoning Key Laboratory of Cosmology and Astrophysics, College of Sciences, Northeastern University, Shenyang 110819, China}

\author{Xin Zhang\orcid{0000-0002-6029-1933}}
\thanks{Corresponding author}
\email{zhangxin@neu.edu.cn}
\affiliation{Liaoning Key Laboratory of Cosmology and Astrophysics, College of Sciences, Northeastern University, Shenyang 110819, China}
\affiliation{MOE Key Laboratory of Data Analytics and Optimization for Smart Industry, Northeastern University, Shenyang 110819, China}
\affiliation{National Frontiers Science Center for Industrial Intelligence and Systems Optimization, Northeastern University, Shenyang 110819, China}

\begin{abstract}
Fast radio bursts (FRBs) probe the integrated free-electron density along extragalactic sightlines and can be used to study late-time cosmic evolution. Conventional Bayesian analyses require repeated evaluations of computationally expensive likelihoods, making inference increasingly challenging as FRB catalogs grow. In this work, we develop an interpretable attention-based neural posterior estimator (AttNPE) for rapid posterior inference of the physical parameters from FRB catalogs with measured redshifts. The attention encoder compresses the catalog into a low-dimensional representation, and a conditional masked autoregressive flow maps samples from a base distribution to the seven-dimensional posterior. AttNPE yields posterior distributions in close agreement with Markov chain Monte Carlo (MCMC), recovering the parameter correlations in both simulated and observed FRB catalogs. For the observed catalog, AttNPE generates $5\times10^4$ posterior samples in $0.036$~s after training, compared with $72.8$~h for the reference MCMC run, giving an elapsed-time ratio of $O(10^6)$. The attention-weight analysis shows that intermediate-redshift FRBs ($0.3\leq z<0.7$) account for the largest share of the total attention averaged over simulated catalogs following the observed redshift distribution, consistent with the combined influence of source abundance and redshift-dependent sensitivity. These results demonstrate the potential of interpretable neural posterior estimation for efficient and repeated inference in FRB cosmology.
\end{abstract}

\maketitle

\section{Introduction}
\label{sec:introduction}

Fast radio bursts (FRBs) are millisecond-duration radio transients with diverse temporal, spectral, and polarization properties, providing important probes of compact objects and their coherent emission processes \cite{Lorimer:2007qn,Masui:2015kmb,Ravi:2016kfj,Cordes:2017eug,Cordes:2019cmq,Petroff:2019tty,Cho:2020gtg,Luo:2020kwy,Hilmarsson:2020npc,Anna-Thomas:2022yvr,Nimmo:2024vwq}. The observed population includes both repeating and apparently nonrepeating sources \cite{Spitler:2016dmz,CHIMEFRB:2020bcn,CHIMEFRB:2021srp,Sun:2024huw,Sun:2025cdq,Sun:2026aeh}, while the detection of a bright radio burst from SGR~J1935+2154 has strengthened the observational link between magnetars and FRBs \cite{Bochenek:2020zxn}. Beyond their source physics, FRBs also carry information about the ionized matter along their propagation paths through their dispersion measures (DMs). However, the observed DM contains contributions from the Milky Way, the host galaxy, and the local source environment in addition to the intergalactic medium (IGM), complicating the extraction of the IGM signal \cite{Bhardwaj:2021hgc,Bhardwaj:2021xaa,Fong:2021xxj,Tendulkar:2020npy,Bhandari:2021pvj,Kirsten:2021llv,KochOcker:2022ook,Niu:2021bnl,Wang:2022mux,Bhandari:2022ton,Caleb:2023atr,Driessen:2023lxj,Gordon:2023cgw,Mckinven:2022fxn,Ibik:2023ugl,Pearlman:2023brm,Shah:2024ywp,Horowicz:2025rhc}. Precise localizations have enabled host-galaxy identification and redshift measurements for an increasing number of FRBs \cite{Chatterjee:2017dqg,Bannister:2019iju,DeepSynopticArrayTeam:2023suu}. With these redshifts and appropriate modeling of the non-IGM contributions, the DM--redshift relation provides a means of probing the ionized IGM and the cosmic expansion history \cite{Ioka:2003fr,Inoue:2003ga,McQuinn:2013tmc,Chatterjee:2017dqg,Marcote:2017wan,Tendulkar:2017vuq,Bannister:2019iju,Ravi:2019alc,Bhandari:2020cde,Marcote:2020ljw,DeepSynopticArrayTeam:2023suu}.

The DM--redshift relation has enabled a census of diffuse cosmic baryons and constraints on the late-time expansion history \cite{Walters:2017afr,Li:2019klc,Macquart:2020lln,Zhao:2020ole,Wei:2019uhh,Qiang:2020vta,Li:2020qei,Dai:2021czy,Hagstotz:2021jzu,James:2022dcx,Yang:2022ftm,Wang:2022ami,Wei:2023avr,2023SCPMA..6620431D,Connor:2024mjg,Acharya:2025ubt,Gao:2025fcr,Xu:2025ddk,Zhuge:2025urk,Zhang:2025yhi,Liu:2025fdf,Sang:2025mti,Sales:2026smc,Liu:2026epv}, complementing other cosmological distance probes \cite{Wang:2019tto,Jin:2021pcv,Wu:2022dgy,Jin:2022qnj,Wang:2025ugc,Wu:2026mox,Zhang:2026axl}. FRBs have also been used to constrain dark-energy dynamics \cite{Deng:2013aga,Gao:2014iva,Yang:2016zbm,Qiu:2021cww,Zhang:2023gye} and dark-sector interactions \cite{Zhao:2022bpd}. Further applications include gravitational-lensing and multimessenger cosmological tests \cite{Li:2017mek,Zhang:2024rra,CHIMEFRB:2022xzl,Leung:2022vcx,LIGOScientific:2022jpr,Moroianu:2022ccs}. Together, these studies establish FRBs as valuable probes of late-time cosmology \cite{Li:2026asg,Cao:2026mhu,Zhang:2026zbs}. For recent reviews of FRB physics and cosmological applications, see Refs.~\cite{Zhang:2022uzl,Wu:2024iyu,Bao:2025esa,Wang:2026tgi}.

However, constructing FRB likelihoods requires careful treatment of the different contributions to the observed DM. The Milky Way halo contribution remains uncertain, while the extragalactic DM varies with the IGM matter distribution and the properties of host galaxies and their environments. These contributions are commonly described by probability distributions and must be convolved and marginalized over when constructing the likelihood for each FRB \cite{2019MNRAS.485..648P,Macquart:2020lln,James:2021jbo,James:2022dcx,Wang:2022ami,Zhang:2025yhi,Zhang:2020mgq,Zhang:2023pqs,Baptista:2023uqu,Liu:2025fdf,Sales:2026smc}. Moreover, survey selection effects and spatial correlations induced by large-scale structure can require more general simulation-based treatments in large-sample analyses \cite{Sharma:2025lql,Reischke:2023blu}. Consequently, each likelihood evaluation becomes more computationally expensive, making Markov chain Monte Carlo (MCMC) inference increasingly time-consuming.

Deep-learning methods provide a promising way to alleviate the computational burden of conventional Bayesian inference and have been widely applied to accelerate data analysis and parameter estimation in astronomy and cosmology \cite{Hezaveh:2017sht,George:2016hay,Gabbard:2017lja,Connor:2018wfr,Zhang:2018jux,Fluri:2019qtp,Wang:2023lif,Fortunato:2024hfm,Wang:2024oei,Sun:2025ypd,Sun:2026wep,Sun:2026kig}. In particular, neural posterior estimation (NPE) learns a conditional approximation to the posterior distribution from simulated parameter--data pairs, enabling rapid posterior sampling after training without repeatedly evaluating the full likelihood \cite{DBLP:conf/nips/DurkanB0P19,Cranmer:2019eaq,Papamakarios:2019fms}. 

Normalizing flows provide flexible conditional density models for correlated and non-Gaussian posteriors and have been widely used in simulation-based inference for cosmology and gravitational-wave astronomy \cite{Alsing:2019xrx,Dax:2021tsq,Green:2020dnx,Dax:2022pxd,Sun:2023vlq,Xiong:2024gpx,Sun:2024ywb,Qin:2025mvj,Sun:2026gsq,Zhang:2026okw,Liu:2026nhu,Sun:2026dga}. Despite these advances, the black-box nature of neural networks can obscure how individual FRBs influence the inferred parameter constraints. Attention-based pooling assigns explicit weights to individual sources when constructing the catalog embedding \cite{DBLP:conf/icml/LeeLKKCT19,Iwanaga:2025muq}. These weights are not necessarily faithful explanations of the inferred posterior \cite{DBLP:conf/naacl/JainW19}. We therefore examine them together with the posterior response to input perturbations and the redshift dependence expected from the DM model.

Previous work applied truncated sequential NPE to FRB cosmology, focusing on  the amplitude of the mean DM--redshift relation and two host-galaxy parameters \cite{Konar:2024mrz}. For cosmological applications, the effects of cosmic expansion, the diffuse baryon content, and the IGM and host-galaxy contributions are coupled in the observed DM--redshift relation. Joint inference is therefore important for accounting for their parameter degeneracies and propagating astrophysical uncertainties into cosmological constraints.

In this work, we develop an attention-based neural posterior estimator (AttNPE) for joint inference of seven cosmological and astrophysical parameters from localized FRB catalogs. The estimator combines an attention encoder with a conditional masked autoregressive flow (MAF), yielding posterior distributions in close agreement with MCMC for simulated and observed catalogs. Our attention and sensitivity analyses show a stronger response to individual high-redshift FRBs, consistent with the increasing IGM contribution to the dispersion measure. When the observed source distribution is taken into account, both analyses instead identify $0.3\leq z<0.7$ as the most influential redshift range for the current catalog.

The organization of this paper is as follows. Section~\ref{sec:method} describes the FRB likelihood, simulation set, neural architecture, and validation protocol. Section~\ref{sec:results} presents the results and discussion. Section~\ref{sec:conclusion} summarizes the conclusions.

\section{Method}
\label{sec:method}

\subsection{FRB dispersion model and conditional likelihood}
\label{subsec:likelihood}

For a localized FRB at redshift $z_i$, the observed dispersion measure, $\DM_{{\rm obs},i}$, is inferred from the frequency-dependent arrival-time delay and represents the integrated free-electron column density along the line of sight. After subtracting the Milky Way interstellar medium (ISM) contribution, $\DM_{{\rm ISM},i}$, predicted by the NE2001 model \citep{Cordes:2002wz}, we define the extragalactic DM excess as
\begin{equation}
\begin{aligned}
\DM_{{\rm E},i}
&\equiv \DM_{{\rm obs},i}-\DM_{{\rm ISM},i}\\
&= \DM_{{\rm halo},i}+\DM_{{\rm IGM},i}
+\frac{\DM_{{\rm host},i}}{1+z_i},
\end{aligned}
\label{eq:dm_decomposition}
\end{equation}
where $\DM_{{\rm halo},i}$, $\DM_{{\rm IGM},i}$, and $\DM_{{\rm host},i}$ denote the contributions from the Milky Way halo, intergalactic medium (IGM), and host environment, respectively. The host contribution is defined in the source rest frame, hence the factor $(1+z_i)^{-1}$. Following the localized-FRB modeling framework in Refs.~\cite{Macquart:2020lln,James:2021jbo,Zhang:2025yhi}, we jointly infer seven parameters,
\begin{equation}
 \bm\theta=\{\Om,\Ob,H_0,\fd,F,\mu_{\rm host},\sigma_{\rm host}\}.
 \label{eq:theta}
\end{equation}
Here $\Om$ is the present-day matter density parameter, $\Ob$ is the present-day baryon density parameter, and $H_0$ is the Hubble constant. Together they specify a flat $\Lambda$CDM background. The parameter $\fd$ is the fraction of baryons in diffuse ionized gas. The parameter $F$ controls the line-of-sight IGM scatter. 

The mean IGM contribution follows \cite{Ioka:2003fr,Inoue:2003ga,Macquart:2020lln}
\begin{equation}
 \left\langle\DMIGM(z)\right\rangle
 =\frac{3cH_0\Ob\fd}{8\pi Gm_p}
 \int_0^z\frac{\chi_e(z')(1+z')}{E(z')}\,\mathrm{d}z',
 \label{eq:mean_dm}
\end{equation}
where $E(z)=H(z)/H_0$, $c$ is the speed of light, $G$ is the gravitational constant, and $m_p$ is the proton mass. We adopt $\chi_e=0.875$, corresponding to the number of free electrons per baryon for a fully ionized
H/He mixture with primordial helium abundance. Because $H_0$, $\Ob$, and $\fd$ enter the mean IGM DM in Eq.~(\ref{eq:mean_dm}) only through their product, we show the posterior for $\mathcal{A}\equiv H_0\Ob\fd$ rather than the individual posteriors for these degenerate parameters. The single-sightline IGM probability density is expressed in terms of $\Delta=\DMIGM/\langle\DMIGM\rangle$ as \cite{McQuinn:2013tmc,Macquart:2020lln}
\begin{equation}
 p_{\rm IGM}(\Delta|z,F)
 =A\Delta^{-\beta}
 \exp\left[-\frac{(\Delta^{-\alpha}-C_0)^2}{2\alpha^2\sigma_{\rm DM}^2}\right],
 \label{eq:igm_pdf}
\end{equation}
with $\alpha=\beta=3$ and $\sigma_{\rm DM}=F/\sqrt{z}$. The constants $A$ and $C_0$ are chosen so that the distribution is normalized and has unit mean.

Following the commonly used host-DM parameterization of Ref.~\cite{Macquart:2020lln}, the host contribution follows a lognormal distribution, with $\ln(\DM_{\rm host}/\DM_0)$ normally distributed, where $\DM_0=1\,\mathrm{pc\,cm^{-3}}$,
\begin{equation}
 \begin{aligned}
 p_{\rm host}(\DM_{\rm host})&=\frac{1}{\sqrt{2\pi}\sigma_{\rm host}\DM_{\rm host}}\\[-1pt]
 &\quad\times\exp\!\left[-\frac{\{\ln(\DM_{\rm host}/\DM_0)-\mu_{\rm host}\}^2}{2\sigma_{\rm host}^2}\right],
 \end{aligned}
 \label{eq:host_pdf}
\end{equation}
with median $\exp(\mu_{\rm host})\,\mathrm{pc\,cm^{-3}}$ and logarithmic width $\sigma_{\rm host}$. Its observed contribution is divided by $1+z$.

Following Ref.~\cite{Zhang:2025yhi}, the Milky Way halo contribution is modeled by a Gaussian with mean $\mu_h=65\,\mathrm{pc\,cm^{-3}}$ and standard deviation $\sigma_h=15\,\mathrm{pc\,cm^{-3}}$. This Gaussian is truncated to the range from $D_{\min}=30\,\mathrm{pc\,cm^{-3}}$ to $D_{\max}=100\,\mathrm{pc\,cm^{-3}}$,
\begin{equation}
\begin{aligned}
p_{\rm halo}(D)&=\frac{\phi[(D-\mu_h)/\sigma_h]}{\sigma_h Z_h}\,\mathbf{1}_{[D_{\min},D_{\max}]}(D),\\
Z_h&=\Phi\!\left(\frac{D_{\max}-\mu_h}{\sigma_h}\right)-\Phi\!\left(\frac{D_{\min}-\mu_h}{\sigma_h}\right),
\end{aligned}
\label{eq:halo_pdf}
\end{equation}
where $\phi$ and $\Phi$ are the standard-normal density and cumulative distribution function, respectively.

The likelihood for a single source is obtained by marginalizing over the host and halo DM contributions,
\begin{equation}
 p_i(\mathrm{DM}_{{\rm E},i}|z_i,\bm\theta)
 =\int \mathrm{d}\DM_{\rm halo}\int \mathrm{d}\DM_{\rm host}\,
 p_{\rm halo}p_{\rm host}p_{\rm IGM},
 \label{eq:single_probability}
\end{equation}
where the IGM argument is fixed by Eq.~(\ref{eq:dm_decomposition}). In terms of the physical IGM contribution, the density entering this convolution is $p_{\rm IGM}(\DM_{\rm IGM}|z_i,\bm\theta)=p_{\rm IGM}(\Delta|z_i,F)/\langle\DMIGM(z_i)\rangle$. Within the localized-FRB modeling framework described above, we treat the measured source redshifts as fixed inputs. Neglecting selection effects associated with detection, host identification, and the DM cuts in Section~\ref{subsec:data}, the catalog likelihood is
\begin{equation}
 \mathcal{L}_{\rm cond}(\bm\theta)
 =\prod_{i=1}^{N_{\rm FRB}}p_i(\mathrm{DM}_{{\rm E},i}|z_i,\bm\theta).
 \label{eq:conditional_likelihood}
\end{equation}

\subsection{FRB sample and mock catalogs}
\label{subsec:data}

We use the localized-FRB compilation of Ref.~\cite{Zhang:2026axl}. The compilation includes localized sources reported in Refs.~\cite{Chatterjee:2017dqg,Marcote:2017wan,Tendulkar:2017vuq,Bannister:2019iju,Ravi:2019alc,Bhandari:2020cde,Marcote:2020ljw,DeepSynopticArrayTeam:2023suu,Connor:2024mjg}.

The initial sample contains 139 localized FRBs, for which source redshifts, observed DMs, and Milky Way disk contributions estimated using the NE2001 model are available. For each FRB, the DM excess entering the likelihood can be rewritten as
\begin{equation}
 \DM_{{\rm E},i}=\DM_{{\rm obs},i}-\DM_{{\rm ISM},i}.
 \label{eq:dme_definition}
\end{equation}
Following the conditional catalog analysis of Ref.~\cite{Zhang:2025yhi}, we treat the observed DM and the Milky Way disk subtraction as fixed inputs. The Milky Way halo contribution is marginalized through $p_{\rm halo}$ in Eq.~(\ref{eq:halo_pdf}). We require finite redshift, observed DM, and Milky Way contribution, and impose
\begin{equation}
\DME>80\,\mathrm{pc\,cm^{-3}},\qquad
\DM_{\rm ISM}\leq0.4~\DM_{\rm obs}.
 \label{eq:quality_cuts}
\end{equation}

The final sample contains 120 FRBs, following the selection and exclusions of Ref.~\cite{Zhang:2026axl}. For the redshift-dependent diagnostics, we divide the sample into four bins: $0\leq z<0.1$, $0.1\leq z<0.3$, $0.3\leq z<0.7$, and $z\geq0.7$, containing 24, 50, 40, and 6 sources, respectively. The highest-redshift source is FRB~20240304B at $z=2.148$ \cite{Caleb:2025uzd}.

All simulated catalogs use the same 120 source redshifts as the observed catalog, with source index $i$ corresponding to the same redshift $z_i$ throughout. The training and validation therefore apply to catalogs with this fixed set of source redshifts. For each parameter vector, the forward simulator draws the IGM, host, and halo DM contributions for each source and combines them using Eq.~(\ref{eq:dm_decomposition}), generating mock catalogs in batches. The mean IGM DM of Eq.~(\ref{eq:mean_dm}) is evaluated by Gauss-Legendre quadrature, and the single-sightline IGM distribution is sampled from a tabulated inverse cumulative distribution. The reference MCMC calculation evaluates the convolved probability density in Eq.~(\ref{eq:single_probability}) for every source at each proposed parameter point. We checked the mean IGM DM computed using Gauss-Legendre quadrature against direct numerical integration over a grid of parameter values and catalog redshifts, finding close agreement.

At each training epoch we regenerate a fresh set of $10^6$ catalogs, reducing repeated exposure to the same realizations and mitigating overfitting to a finite training set. We additionally generate two independent sets of $10^5$ catalogs for validation and testing, respectively. The parameter priors are listed in Table~\ref{tab:priors}.

\begin{table}[t]
\caption{Prior ranges for the seven parameters used in the simulator. All priors are independent, with a uniform prior on $e^{\mu_{\rm host}}$ and uniform priors on the remaining parameters over their listed intervals.}
\label{tab:priors}
\begin{ruledtabular}
\begin{tabular}{lc}
Parameter & Prior \\
\hline
$\Om$ & $[0.28,0.35]$ \\
$\Ob$ & $[0.01,0.10]$ \\
$H_0$ & $[20,140]\,\mathrm{km\,s^{-1}\,Mpc^{-1}}$ \\
$\fd$ & $[0.84,0.96]$ \\
$F$ & $[0.011,0.5]$ \\
$\mu_{\rm host}$ & $[\ln 20,\ln 200]$ \\
$\sigma_{\rm host}$ & $[0.2,2.0]$ \\
\end{tabular}
\end{ruledtabular}
\end{table}

\subsection{Attention-based catalog embedding}
\label{subsec:attnpe}

The neural estimator takes a catalog $\bm x=\{(z_i,\DM_{{\rm E},i})\}_{i=1}^{120}$ as input, consisting of the measured redshift and excess DM of each source. For the $i$th FRB, the input features are
\begin{align}
 \bm u_i={}&\left[
 \ell_i,\ z_i,\ \ln(1+z_i),
 \frac{\ell_i}{\sqrt{\max(z_i,10^{-4})}},
 \ell_i-\overline{\ell}
 \right],\nonumber\\
\ell_i={}&\ln\left[1+\frac{\mathrm{DM}_{{\rm E},i}}{1\,\mathrm{pc\,cm^{-3}}}\right],
\label{eq:token}
\end{align}
where $\overline{\ell}=N_{\rm FRB}^{-1}\sum_i\ell_i$ is the catalog mean of the logarithmically transformed excess DM. These features provide dimensionless inputs to the neural network.

\begin{figure*}[!t]
\centering
\includegraphics[width=0.82\textwidth]{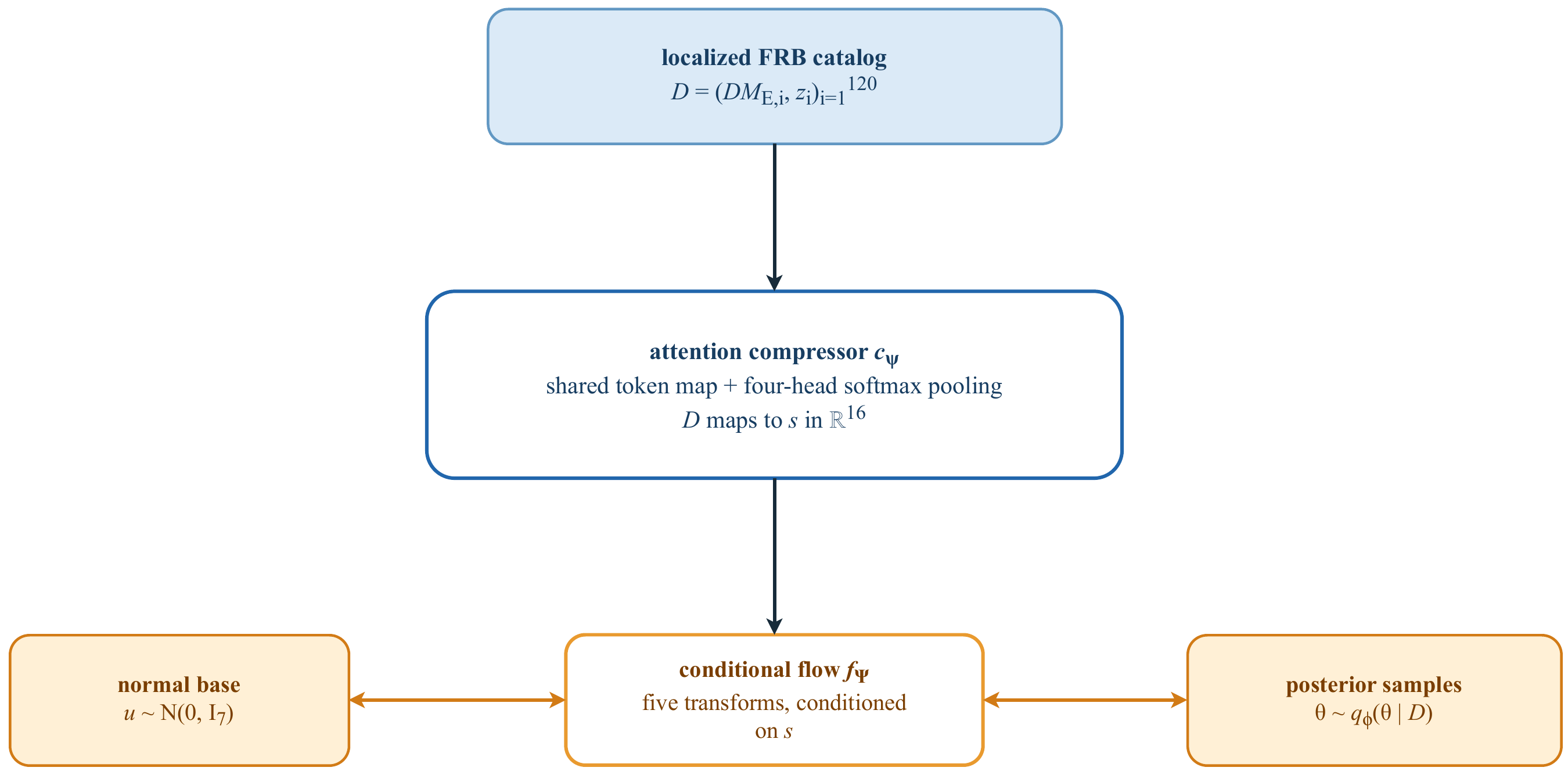}
\caption{AttNPE neural posterior architecture. Each localized FRB is mapped to a 64-dimensional source embedding. Four attention heads form permutation-invariant summaries that are compressed to the context vector $\bm{s}$, and a five-transform conditional MAF maps a seven-dimensional base draw to the joint posterior.}
\label{fig:architecture}
\end{figure*}

A shared encoder maps each source record to a 64-dimensional representation $\bm h_i$. The source encoder uses linear layers of widths $128$ and $64$ with Gaussian error linear unit (GELU) activations \cite{Hendrycks:2016qxa}, layer normalization, and dropout $0.02$. Four attention heads use a $64\rightarrow64\rightarrow4$ multilayer perceptron to assign weights
\begin{equation}
 a_{ki}=\frac{\exp[g_k(\bm h_i)]}{\sum_j\exp[g_k(\bm h_j)]},
 \label{eq:attention}
\end{equation}
where $g_k(\bm h_i)$ is the score assigned to source $i$ by attention head $k$. The pooled summaries are $\bm s_k=\sum_i a_{ki}\bm h_i$. This pooling is invariant to the order of the source records \cite{DBLP:conf/nips/ZaheerKRPSS17,DBLP:conf/icml/LeeLKKCT19}. Each of the four attention heads produces a 64-dimensional summary vector of the full catalog. These vectors are then combined and compressed into a 16-dimensional representation $\bm s$, which provides the conditioning input to the posterior estimator. The complete encoder-flow architecture is shown in Fig.~\ref{fig:architecture}.

\subsection{Conditional normalizing-flow posterior}
\label{subsec:flow}

AttNPE approximates the posterior $p(\bm\theta|\bm x)$ with a conditional density $q_\phi(\bm\theta|\bm s)$, where $\bm s$ is the representation produced by the attention encoder. We also write this density as $q_\phi(\bm\theta|\bm x)$ when referring to the complete encoder--flow model. Before flow evaluation, each physical parameter is linearly rescaled using its prior bounds. A conditional MAF with five affine autoregressive transforms maps a seven-dimensional standard normal  base distribution to the posterior in these normalized coordinates \cite{DBLP:conf/nips/PapamakariosMP17}. Each transform uses three hidden layers of width 128, with the parameter order alternating between successive transforms.

The estimator is implemented with \texttt{PyTorch} \cite{DBLP:conf/nips/PaszkeGMLBCKLGA19} and \texttt{Lampe}. The MCMC cross-check uses the \texttt{emcee} ensemble sampler \cite{Foreman-Mackey:2012any}, with \texttt{Bilby} \cite{Ashton:2018jfp} used for sampler setup and posterior handling.

\begin{table}[t]
\caption{Architecture and optimization settings of AttNPE.}
\label{tab:model_settings}
\begin{ruledtabular}
\begin{tabular}{lc}
Setting & Value \\
\hline
Source embedding dimension & $64$ \\
Catalog embedding dimension & $16$ \\
Number of attention heads & $4$ \\
Number of MAF transformations & $5$ \\
Source-encoder hidden-layer widths & $128 \rightarrow 64$ \\
Hidden layers per MAF transform & $3$ \\
Hidden-layer width per MAF transform & $128$ \\
Batch size & $2048$ \\
Learning rate & $2\times10^{-4}$ \\
Weight decay & $10^{-4}$ \\
Dropout rate & $0.02$ \\
Attention smoothness coefficient $\lambda_{\rm sm}$ & $10^{-4}$ \\
\end{tabular}
\end{ruledtabular}
\end{table}

The encoder and flow are optimized jointly by minimizing the conditional negative log density, with a weak smoothness regularization on the attention weights,
\begin{align}
 \mathcal{L}_{\rm train}={}&-\frac{1}{B}\sum_{b=1}^{B}
 \ln q_\phi(\bm\theta_b|\bm s_b)\nonumber\\
  &+\frac{\lambda_{\rm sm}}{BKN}
 \sum_{b=1}^{B}\sum_{k=1}^{K}\sum_{i=1}^{N-1}
 (a_{bk,i+1}-a_{bki})^2,
\label{eq:loss}
\end{align}
where $B$, $K$, and $N$ are the batch size, number of attention heads, and number of sources, respectively, and $\lambda_{\rm sm}=10^{-4}$. Before this penalty is evaluated, source records are ordered by increasing redshift. The regularizer therefore favors attention weights that vary smoothly between neighboring redshift entries, without imposing an increasing or decreasing trend. The learned weights are interpreted in the context of this smoothness constraint. The pooling operation itself remains permutation invariant at inference. The architecture and optimization settings are summarized in Table~\ref{tab:model_settings}.

\subsection{Posterior validation and interpretability diagnostics}
\label{subsec:diagnostics}

We assess the learned posterior using marginal probability-probability (PP) curves and the joint Tests of Accuracy with Random Points (TARP) diagnostic \cite{DBLP:conf/icml/LemosCHL23}. We evaluate marginal PP curves and joint TARP coverage using the same 500 test catalogs. For parameter $\theta_d$ in catalog $i$, the posterior percentile rank is
\begin{equation}
 u_{id}=\int_{-\infty}^{\theta_{id}^{\rm true}}
 q_{id}(\theta'_d|\bm x_i)\,\mathrm{d}\theta'_d.
 \label{eq:rank}
\end{equation}
Here $q_{id}$ is the marginal posterior density for parameter $\theta_d$ in catalog $i$. We estimate $u_{id}$ as the fraction of posterior samples below the injected value $\theta_{id}^{\rm true}$. For each parameter, the PP curve shows the cumulative distribution of these percentile ranks across the simulated catalogs. For calibrated posteriors, the ranks should be uniformly distributed on $[0,1]$, so the PP curve should follow the diagonal. Following the PP-plot procedure implemented in \texttt{Bilby}, we apply a Kolmogorov--Smirnov (KS) test against the uniform distribution to the percentile ranks of each parameter \cite{Ashton:2018jfp,Dax:2021tsq}. We report the seven marginal $p$-values separately because the ranks for different parameters can be correlated. TARP separately tests joint coverage after mapping the parameters and posterior samples to the unit prior hypercube. We compare the mean TARP coverage curve, computed using Euclidean distances, with the diagonal.

We also compare AttNPE with MCMC posteriors obtained from the same likelihood and priors using the \texttt{emcee} ensemble sampler \cite{Foreman-Mackey:2012any}. We perform these comparisons for a representative simulated catalog and the observed FRB catalog. For one-dimensional marginal densities $p$ and $q$, we use the Jensen-Shannon divergence (JSD)
\begin{equation}
 D_{\rm JS}(p,q)=\frac{1}{2}D_{\rm KL}(p\Vert \frac{p+q}{2})+\frac{1}{2}D_{\rm KL}(q\Vert \frac{p+q}{2}),
 \label{eq:jsd}
\end{equation}
where $D_{\rm KL}$ denotes the Kullback-Leibler divergence and $D_{\rm JS}=0$ for identical distributions. We compute it from normalized posterior histograms and also report the ratio
\begin{equation}
r_{68}=\frac{w_{68}^{\rm AttNPE}}{w_{68}^{\rm MCMC}},
\qquad w_{68}=q_{84}-q_{16}.
\label{eq:width_ratio}
\end{equation}
Here $q_{16}$ and $q_{84}$ are the 16th and 84th posterior percentiles. A value of $r_{68}$ close to unity indicates agreement between the central 68\% marginal interval widths from the two methods.

To examine which redshift ranges are used by the catalog encoder, we combine the attention weights with a single-source DM replacement. The attention weights describe how the encoder weights individual source records, whereas the DM replacement measures the change in the inferred posterior after perturbing one source. We average the attention weights at each fixed source index over the four heads and the held-out simulated catalogs, which share the redshifts of the observed FRB catalog. We then compare these quantities with the redshift dependence of the mean IGM DM. For held-out catalogs, we replace the DM of one source with the value at the same source index in a randomly selected donor catalog while keeping its redshift fixed~\cite{DBLP:journals/ml/Breiman01,DBLP:journals/jmlr/FisherRD19}. This replacement preserves the one-source marginal distribution at fixed redshift while breaking its association with the parameters $\bm\theta_n$ that generated the target catalog $\bm x_n$. It also changes the dependence between that source and the remaining catalog induced by their shared parameters. We use this perturbation to measure the estimator's sensitivity to the source DM. For the modified catalog $\bm x_n^{(i)}$, we evaluate the change in the negative log posterior density at the generating parameters $\bm\theta_n$, denoted by $\Delta\mathrm{NLL}_i$,
\begin{equation}
\Delta\mathrm{NLL}_i=
\mathbb{E}_n\left[
-\ln q_\phi(\bm\theta_n|\bm x_n^{(i)})
+\ln q_\phi(\bm\theta_n|\bm x_n)
\right] .
\label{eq:permutation_nll}
\end{equation}
For each source, $\Delta\mathrm{NLL}_i$ is averaged over $5\times10^4$ held-out catalogs and evaluated with three independent donor draws. Positive values indicate that replacing the source DM lowers the density assigned to the generating parameters.

For each inferred parameter, let $\bar\theta_{nd}$ and $w_{nd}=q_{84,nd}-q_{16,nd}$ denote the posterior mean and central 68\% width for catalog $n$, where $q_{84,nd}$ and $q_{16,nd}$ are the 84th and 16th posterior percentiles. We define the posterior-mean displacement normalized by the unperturbed central 68\% width,
\begin{equation}
 R_{id}=\left\langle
 \frac{\left|\bar\theta_{nd}(\bm x_n^{(i)})-
 \bar\theta_{nd}(\bm x_n)\right|}{w_{nd}(\bm x_n)}
 \right\rangle_n .
 \label{eq:permutation_shift}
\end{equation}
For each parameter, we sum $R_{id}$ over the sources in each redshift bin and normalize the four bin sums to unity. The resulting fractions measure each bin's share of the total posterior-mean response and depend on both the response of individual sources and the number of sources in the bin.

\section{Results and discussion}
\label{sec:results}

\subsection{Reliability of the inferred posterior}
\label{subsec:calibration}

\begin{figure}[t]
\centering
\includegraphics[width=\columnwidth]{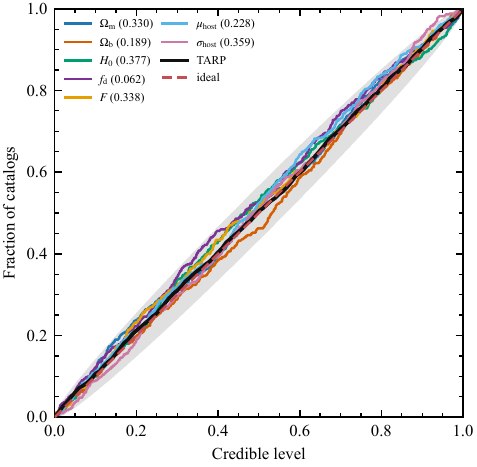}
\caption{Marginal PP curves and joint TARP coverage for 500 test catalogs. Colored curves show the marginal PP curves, with the corresponding KS $p$-values given in parentheses in the legend. The thick black curve shows the mean TARP coverage computed using the Euclidean distance. The red dashed line indicates ideal coverage. The gray shaded region denotes the $3\sigma$ confidence interval for the marginal PP curves.}
\label{fig:pp_calibration}
\end{figure}

We evaluate the calibration of AttNPE on independent simulations using marginal PP curves and joint TARP coverage for the seven-dimensional posterior. Figure~\ref{fig:pp_calibration} shows the resulting marginal and joint coverage curves.

The marginal PP curves and the mean Euclidean TARP coverage curve in Fig.~\ref{fig:pp_calibration} lie close to the diagonal. The individual marginal KS $p$-values, shown in the legend, are all greater than $0.05$, consistent with uniformly distributed posterior percentile ranks. Together, these checks support the reliability of the posterior uncertainty estimates obtained with AttNPE.

\begin{figure*}[t]
\centering
\includegraphics[width=0.82\textwidth]{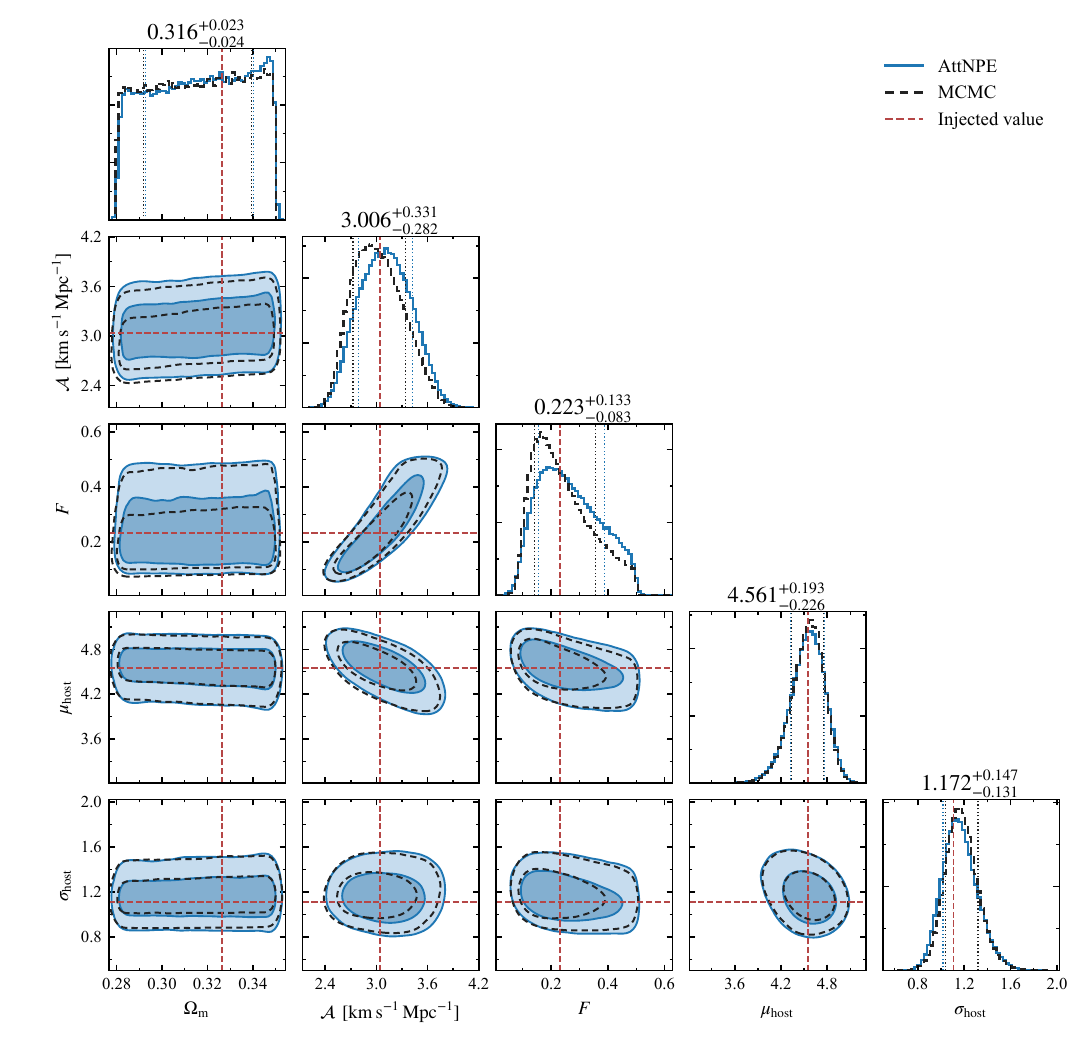}
\caption{AttNPE and MCMC posteriors for a representative simulated catalog. Diagonal panels show the one-dimensional marginals and off-diagonal panels show pairwise parameter correlations. Blue solid curves and shaded contours show AttNPE, black dashed curves and contours show MCMC, and red dashed lines mark the injected parameter values. The values above the diagonal panels give the MCMC posterior medians and central 68\% credible intervals. Contours enclose 68\% and 95\% posterior mass.}
\label{fig:mock_corner}
\end{figure*}

Coverage tests assess calibration across repeated simulated catalogs. We additionally compare AttNPE with a direct MCMC calculation for a representative simulated catalog to examine its posterior shape and parameter correlations. Figure~\ref{fig:mock_corner} shows that all injected values lie within the central 68\% marginal intervals for both methods and that the principal degeneracy directions agree. The parameter summaries and JSD values are listed in Table~\ref{tab:mock_parameters}. The JSD values are below $1\times10^{-2}$ and the interval-width ratios are close to unity. The small JSD values and near-unit width ratios indicate close agreement with MCMC for the representative realization, while the PP curves and marginal KS tests assess posterior calibration across the simulated catalogs.

\begin{table*}[t]
\caption{Posterior summaries for the representative simulated catalog.}
\label{tab:mock_parameters}
\begin{ruledtabular}
\begin{tabular}{lccccc}
Parameter & Injected & AttNPE & MCMC & JSD & $r_{68}$ \\
\hline
$\Omega_{\rm m}$ & 0.3264 & $0.3172^{+0.0230}_{-0.0245}$ & $0.3162^{+0.0233}_{-0.0244}$ & $6.68\times10^{-4}$ & 0.998 \\
$\mathcal{A}\,[\mathrm{km\,s^{-1}\,Mpc^{-1}}]$ & 3.039 & $3.103^{+0.317}_{-0.311}$ & $3.006^{+0.331}_{-0.282}$ & $9.15\times10^{-3}$ & 1.025 \\
$F$ & 0.2321 & $0.2544^{+0.1347}_{-0.0997}$ & $0.2229^{+0.1333}_{-0.0825}$ & $7.77\times10^{-3}$ & 1.086 \\
$\mu_{\rm host}$ & 4.5498 & $4.5566^{+0.2074}_{-0.2337}$ & $4.5605^{+0.1934}_{-0.2262}$ & $5.64\times10^{-4}$ & 1.051 \\
$\sigma_{\rm host}$ & 1.1069 & $1.1567^{+0.1620}_{-0.1398}$ & $1.1716^{+0.1466}_{-0.1309}$ & $2.76\times10^{-3}$ & 1.088 \\
\end{tabular}
\end{ruledtabular}
\end{table*}

\subsection{Redshift dependence of the posterior response}
\label{subsec:physics_information}

To interpret how AttNPE uses FRBs at different redshifts, we compare its attention weights with the posterior response to single-source DM replacement and with the analytic response of the mean IGM DM. These comparisons relate the network's weighting of individual sources to changes in the inferred parameters and to the underlying DM--redshift relation.

\begin{figure*}[t]
\centering
\includegraphics[width=0.98\textwidth]{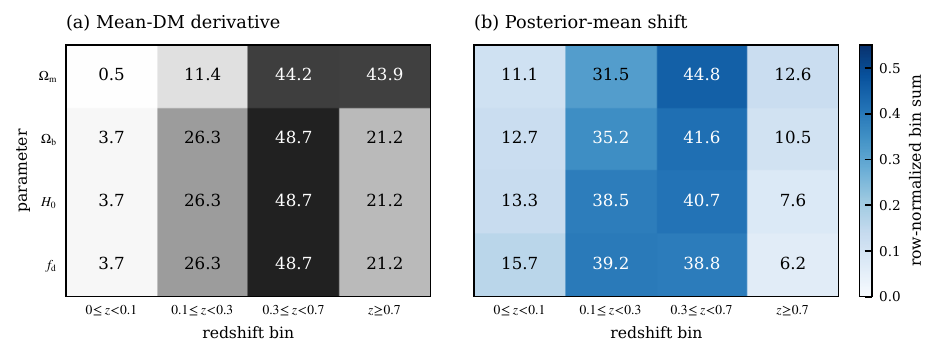}
\caption{Analytic and estimator-based redshift dependences for $\Omega_{\rm m}$, $\Omega_{\rm b}$, $H_0$, and the diffuse-baryon fraction $\fd$. The left matrix gives the redshift-bin sums of $|\partial\langle\DMIGM\rangle/\partial\theta_d|$ calculated from Eq.~(\ref{eq:mean_dm}). The right matrix gives the bin sums of the AttNPE posterior-mean shift $R_{id}$ obtained from held-out simulated catalogs after single-source DM replacement. Each row is normalized over the four bins, and the cell labels give percentages. $H_0$ is in units of $\rm km~s^{-1}~Mpc^{-1}$.}
\label{fig:binned_sensitivity}
\end{figure*}

Figure~\ref{fig:binned_sensitivity} compares two redshift-binned diagnostics for $\Omega_{\rm m}$, $\Omega_{\rm b}$, $H_0$, and $\fd$. The left matrix is constructed by summing the absolute derivatives $|\partial\langle\DMIGM(z_i)\rangle/\partial\theta_d|$ over the sources in each redshift bin. The right matrix contains the corresponding bin sums of the normalized posterior-mean shifts $R_{id}$ induced by single-source DM replacement. Each row is normalized to unity. For $\Omega_{\rm m}$, the four redshift bins contribute 11.1\%, 31.5\%, 44.8\%, and 12.6\% of the normalized summed shift. The $0.3\leq z<0.7$ bin is also the largest for $\Omega_{\rm b}$ and $H_0$, with fractions of 41.6\% and 40.7\%, respectively; for $\fd$, the two intermediate bins contribute nearly equally.

The differences between the two matrices reflect the distinction between the parameter dependence of the mean DM and the response of the full posterior. For $\Omega_{\rm b}$, $H_0$, and $\fd$, the normalized analytic responses are identical because these parameters enter Eq.~(\ref{eq:mean_dm}) through a common multiplicative amplitude. Their posterior responses nevertheless differ because the inferred constraints also depend on IGM and host scatter, parameter priors, and correlations with the remaining parameters. The analytic response therefore provides a physical reference for interpreting the posterior shifts.

\begin{figure*}[t]
\centering
\includegraphics[width=0.98\textwidth]{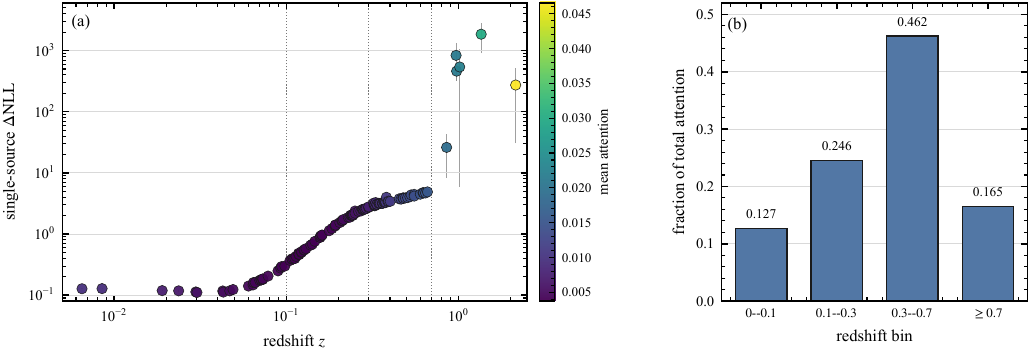}
\caption{Source-level posterior response and redshift-binned attention. (a) Change in the negative log posterior density at the generating parameters, $\Delta\mathrm{NLL}_i$, as a function of source redshift. Point colors indicate attention weights averaged over the four heads and the held-out simulated catalogs, and error bars show the standard deviation across three independent donor draws. (b) Fraction of the total attention obtained by summing source weights within each redshift bin.}
\label{fig:source_dm_replacement}
\end{figure*}

Figure~\ref{fig:source_dm_replacement}(a) shows the single-source response $\Delta\mathrm{NLL}_i$ as a function of redshift, with the learned attention weights indicated by point color. The response generally increases with redshift, consistent with the increasing IGM contribution along longer sightlines. Figure~\ref{fig:source_dm_replacement}(b) shows that the $0.3\leq z<0.7$ bin accounts for 46.2\% of the total mean attention weight, the largest fraction among the four redshift bins. This bin also produces the largest summed posterior-mean shifts for $\Omega_{\rm m}$, $\Omega_{\rm b}$, and $H_0$ in Fig.~\ref{fig:binned_sensitivity}. The diagnostics therefore distinguish the response to an individual FRB from the sum of individual responses within a redshift bin.

Only six sources in the present catalog lie at $z\geq0.7$, limiting the contribution of this population despite its generally larger source-level responses. This interpretation is consistent with forecast studies showing that FRB cosmological constraints depend strongly on the number of localized bursts and on the scatter of the DM-redshift relation \cite{Walters:2017afr,Zhao:2020ole,Qiu:2021cww}. The simulation-based FRB analysis of Ref.~\cite{Konar:2024mrz} likewise emphasizes that enlarging the sample of FRBs with identified host galaxies will strengthen the resulting constraints. Increasing the number of accurately localized FRBs at $z\gtrsim0.7$ may therefore increase the relative contribution of the high-redshift population in future catalogs.

\subsection{Application to the observed FRB catalog}
\label{subsec:real_catalog}

We apply AttNPE to the observed localized-FRB catalog and compare its posterior directly with MCMC.

\begin{figure*}[t]
\centering
\includegraphics[width=0.98\textwidth]{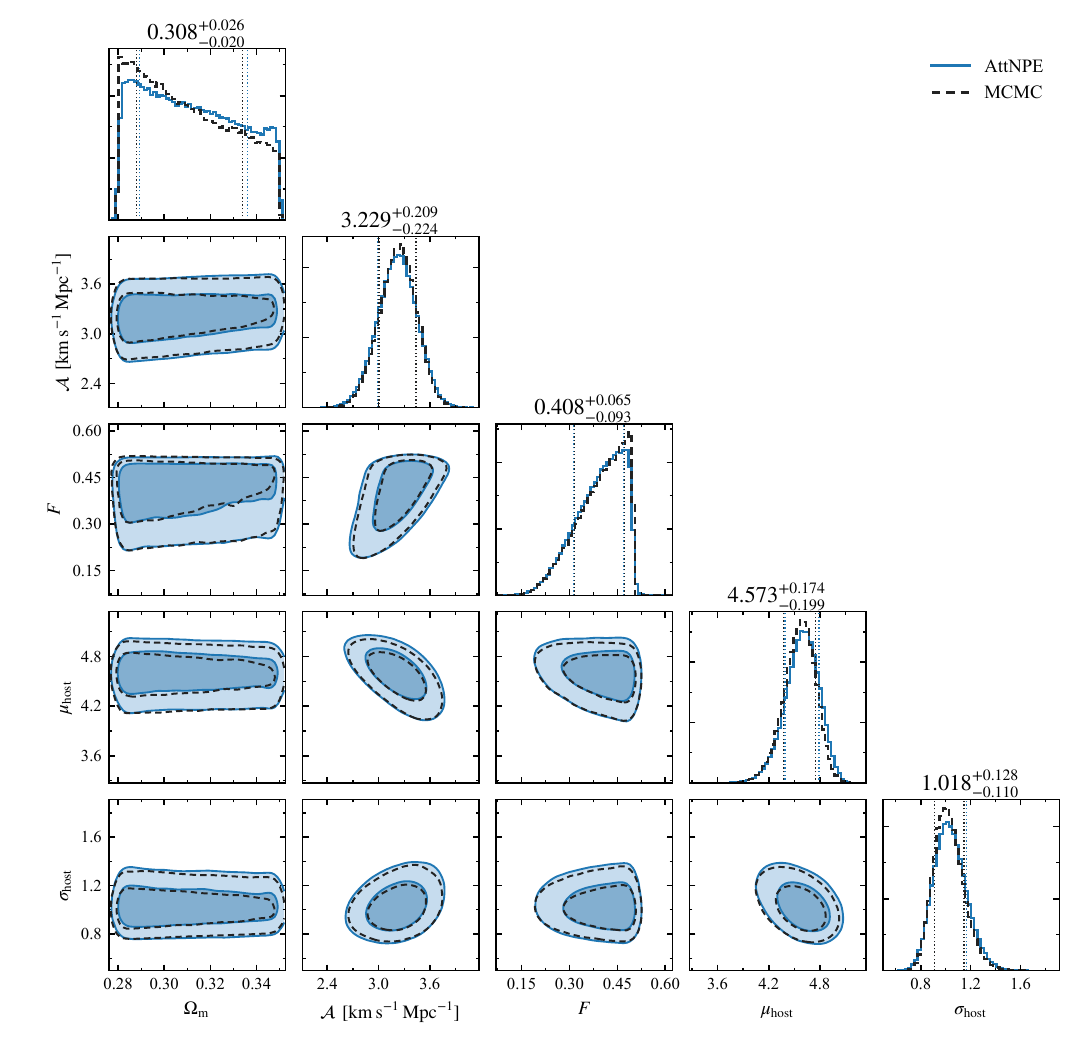}
\caption{AttNPE-MCMC comparison for the observed FRB catalog. Diagonal panels show one-dimensional marginals and off-diagonal panels show pairwise parameter correlations. Blue solid curves and shaded contours show AttNPE, and black dashed curves and contours show MCMC. The values above the diagonal panels give the MCMC posterior medians and central 68\% credible intervals. Contours enclose 68\% and 95\% posterior mass.}
\label{fig:real_corner_overlay}
\end{figure*}

Figure~\ref{fig:real_corner_overlay} compares the AttNPE and MCMC posteriors for the observed catalog, with posterior summaries in Table~\ref{tab:real_parameters}. The marginal JSD is below 0.0043 for all five displayed quantities, and $r_{68}$ ranges from 0.993 to 1.095. In particular, the two methods yield similar constraints on the mean IGM dispersion-measure amplitude $\mathcal{A}$.

\begin{table*}[t]
\caption{Posterior summaries for the observed FRB catalog.}
\label{tab:real_parameters}
\begin{ruledtabular}
\begin{tabular}{lcccc}
Parameter & AttNPE & MCMC & JSD & $r_{68}$ \\
\hline
$\Omega_{\rm m}$ & $0.3107^{+0.0254}_{-0.0215}$ & $0.3077^{+0.0263}_{-0.0198}$ & $2.10\times10^{-3}$ & 1.018 \\
$\mathcal{A}\,[\mathrm{km\,s^{-1}\,Mpc^{-1}}]$ & $3.221^{+0.224}_{-0.232}$ & $3.229^{+0.209}_{-0.224}$ & $8.04\times10^{-4}$ & 1.056 \\
$F$ & $0.4040^{+0.0657}_{-0.0913}$ & $0.4084^{+0.0648}_{-0.0933}$ & $3.61\times10^{-3}$ & 0.993 \\
$\mu_{\rm host}$ & $4.5994^{+0.1887}_{-0.2080}$ & $4.5728^{+0.1741}_{-0.1988}$ & $4.28\times10^{-3}$ & 1.064 \\
$\sigma_{\rm host}$ & $1.0303^{+0.1385}_{-0.1214}$ & $1.0179^{+0.1278}_{-0.1096}$ & $2.42\times10^{-3}$ & 1.095 \\
\end{tabular}
\end{ruledtabular}
\end{table*}

Table~\ref{tab:timing} compares the post-training posterior-sampling times for the observed FRB catalog. AttNPE generates $5\times10^4$ posterior samples in $0.036$~s on an NVIDIA RTX A6000 GPU, whereas the MCMC calculation retains $1.04\times10^6$ samples in $72.8$~h using 28 workers on Intel Xeon w9-3495X hardware. The corresponding elapsed-time ratio is $7.28\times10^6$, and the average time per retained sample is lower for AttNPE by a factor of $3.49\times10^5$. These values characterize post-training inference; catalog simulation and network optimization are performed once and amortized over subsequent posterior evaluations.

\begin{table*}[t]
\caption{Post-training posterior-sampling time for the observed FRB catalog. Both ratios are given relative to AttNPE.}
\label{tab:timing}
\begin{ruledtabular}
\begin{tabular}{lccccc}
Method & Posterior samples & Elapsed time & Elapsed-time ratio & Time per sample & Per-sample ratio \\
\hline
AttNPE & $5.00\times10^4$ & $0.036$ s & -- & $7.20\times10^{-7}$ s & -- \\
MCMC & $1.04\times10^6$ & $72.8$ h & $7.28\times10^6$ & $2.52\times10^{-1}$ s & $3.49\times10^5$ \\
\end{tabular}
\end{ruledtabular}
\end{table*}

\section{Conclusion}
\label{sec:conclusion}

We have developed AttNPE, an attention-based neural posterior estimator, for joint inference of seven cosmological and astrophysical parameters from localized FRB catalogs. Marginal PP and joint TARP tests show coverage consistent with the nominal levels over independent simulations, while direct MCMC comparisons show that AttNPE reproduces the one- and two-dimensional posterior structure for a representative simulated catalog and the observed catalog.

The redshift analysis shows that individual high-redshift FRBs generally produce larger posterior responses, consistent with the increasing IGM contribution along longer sightlines. The six sources at $z\geq0.7$ receive the largest mean attention weight per source, while the 40 sources at $0.3\leq z<0.7$ account for 46.2\% of the total attention averaged over simulated catalogs at the observed redshifts. The intermediate-redshift bin also produces the largest summed posterior-mean shifts for $\Omega_{\rm m}$, $\Omega_{\rm b}$, and $H_0$. The prominence of this bin therefore reflects both the response to individual FRBs and their abundance in the current sample. These findings motivate expanding the sample of accurately localized high-redshift FRBs to better probe the redshift dependence of the DM--redshift relation and strengthen its use in constraining cosmic expansion and the diffuse baryon content.

For the observed catalog, AttNPE generates $5\times10^4$ posterior samples in $0.036$~s after training, compared with $72.8$~h for the reference MCMC run, giving an elapsed-time ratio of $O(10^6)$. Once trained, AttNPE provides rapid posterior sampling and amortizes the simulation and optimization costs over repeated posterior evaluations within the adopted training setup.

The current analysis infers the posterior conditional on measured source redshifts under the adopted dispersion model and priors. Future population-level extensions will incorporate event rates, redshift distributions, survey-selection effects, and more flexible IGM and host-galaxy prescriptions. Inspired by recent work on flexible gravitational-wave parameter estimation \cite{Kofler:2025dux}, we will also explore architectures and training strategies that enable a single trained model to analyze FRB catalogs with different source numbers and redshift coverage. These developments would broaden the application of AttNPE to larger and more heterogeneous localized-FRB samples. The present results demonstrate that AttNPE can reproduce the reference posterior rapidly while revealing how the estimator uses sources at different redshifts.

\section*{Acknowledgements}

This work was supported by the National SKA Program of China (Grants Nos. 2022SKA0110200 and 2022SKA0110203), the National Natural Science Foundation of China (Grants Nos. 12473001, 11975072, 11875102, and 11835009), and the National 111 Project (Grant No. B16009). OpenAI’s ChatGPT and Codex [GPT-6 Astra] were used to improve the language of the manuscript, help cross-check algebraic calculations and derivations, and assist with code modification and numerical checks. The authors take full responsibility for the scientific content and conclusions of this work.

\section*{Data Availability}

The localized-FRB measurements were compiled from the publications cited in Section~\ref{subsec:data}. The catalog used in this analysis, the full simulated training suite, and the data supporting the figures are available from the corresponding author upon reasonable request. The figure data will be deposited in a public repository upon acceptance.

\clearpage
\bibliography{paper}

\end{document}